\documentclass[prb,twocolumn,superscriptaddress,nobibnotes]{revtex4}

\usepackage{graphicx}
\usepackage{dcolumn}
\usepackage{bm}
\usepackage{times,mathptmx}
\usepackage{color}
\usepackage{multirow}
\usepackage{enumitem}
\usepackage{chngpage}
\usepackage{amsmath}

\begin{document}


\title{\bf High-Q two-dimensional lithium niobate photonic crystal slab nanoresonators}

\author{Mingxiao Li}
\affiliation{Department of Electrical and Computer Engineering, University of Rochester, Rochester, NY 14627}
\author{Hanxiao Liang}
\affiliation{Department of Electrical and Computer Engineering, University of Rochester, Rochester, NY 14627}
\author{Rui Luo}
\affiliation{Institute of Optics, University of Rochester, Rochester, NY 14627}
\author{Yang He}
\affiliation{Department of Electrical and Computer Engineering, University of Rochester, Rochester, NY 14627}
\author{Qiang Lin}
\email{qiang.lin@rochester.edu}
\affiliation{Department of Electrical and Computer Engineering, University of Rochester, Rochester, NY 14627}
\affiliation{Institute of Optics, University of Rochester, Rochester, NY 14627}



\begin{abstract}
Lithium niobate (LN), known as "silicon of photonics," exhibits outstanding material characteristics with great potential for broad applications. Enhancing light-matter interaction in the nanoscopic scale would result in intriguing device characteristics that enable revealing new physical phenomena and realizing novel functionalities inaccessible by conventional means. High-Q two dimensional (2D) photonic crystal (PhC) slab nanoresonators are particularly suitable for this purpose, which, however, remains open challenge to be realized on the lithium niobate platform. Here we take an important step towards this direction, demonstrating 2D LN PhC slab nanoresonators with optical Q as high as $3.51 \times 10^5$, about three orders of magnitude higher than other 2D LN PhC structures reported to date. The high optical quality, tight mode confinement, together with pure polarization characteristics of the devices enable us to reveal peculiar anisotropy of photorefraction quenching and unique anisotropic thermo-optic nonlinear response, which have never been reported before. They also allow us to observe third harmonic generation for the first time in on-chip LN nanophotonic devices, and strong orientation-dependent generation of second harmonic. The demonstrated high-Q 2D LN PhC nanoresonators not only offer an excellent device platform for the exploration of extreme nonlinear and quantum optics at single-photon and few-photon level, but also open up a great avenue towards future development of large-scale integrated LN photonic circuits for energy efficient nonlinear photonic and electro-optic signal processing.

\end{abstract}

\maketitle

\section*{Introduction}

Photonic crystal (PhC) nanoresonators exhibit exceptional capability of controlling light confinement and light-matter interactions in the sub-wavelength scale, which forms a crucial foundation for many applications such as signal processing \cite{Notomi10}, information storage \cite{Notomi14}, bio-sensing \cite{Fan08}, nonlinear photonics \cite{Soljacic04}, cavity quantum electrodynamics \cite{Lodahl15}, among many others. Among various photonic crystal structures, two-dimensional (2D) photonic crystal slabs exhibit significant advantage in the engineering of the density of photonic states, the flexibility of device structure design, the scalability of optoelectronic integration, and the compatibility with current nanofabrication technology. These excellent characteristics have excited tremendous interest in recent years to develop 2D PhC slab nanoresonators on a variety of material platforms \cite{Notomi10, Notomi14, Fan08, Soljacic04, Lodahl15, Noda05, Noda07, Benisty08, Eggleton11, Trivino14, Debnath17}.

Lithium niobate (LN), known as ``silicon of photonics," \cite{Buse09} exhibits outstanding electro-optic, nonlinear optical, acousto-optic, piezoelectric, photorefractive, pyroelectric, and photoconductive properties \cite{Gaylord85}, promising for broad applications \cite{Arizmendi04}. The great application potential has attracted significant attention recently to develop LN photonic devices on chip-scale platforms \cite{Gunter07, Fathpour13, Reano14, Loncar14, Cheng15, Pertsch15, Xu15, Hu15, Shayan16, Jiang16, Bower16, Fathpour17, Loncar17, Luo17, Sun17, Luo172, Amir17, Liang17, Buse17, Loncar173, Peruzzo17, Loncar172}. However, realizing high-quality 2D LN PhC structures remains significant challenge \cite{Baida05, Gu06, Salut06, Gunter09, Pertsch10, Laude10, Baida11, Bernal12, Diziain13, Wang14, Pertsch14}, which becomes the major obstacle hindering the exploration of optical phenomena in the nanoscopic scale that would potentially result in intriguing device characteristics and novel functionalities inaccessible by conventional means.

\begin{figure*}[hbtp]
    \centering\includegraphics[width=2.0\columnwidth]{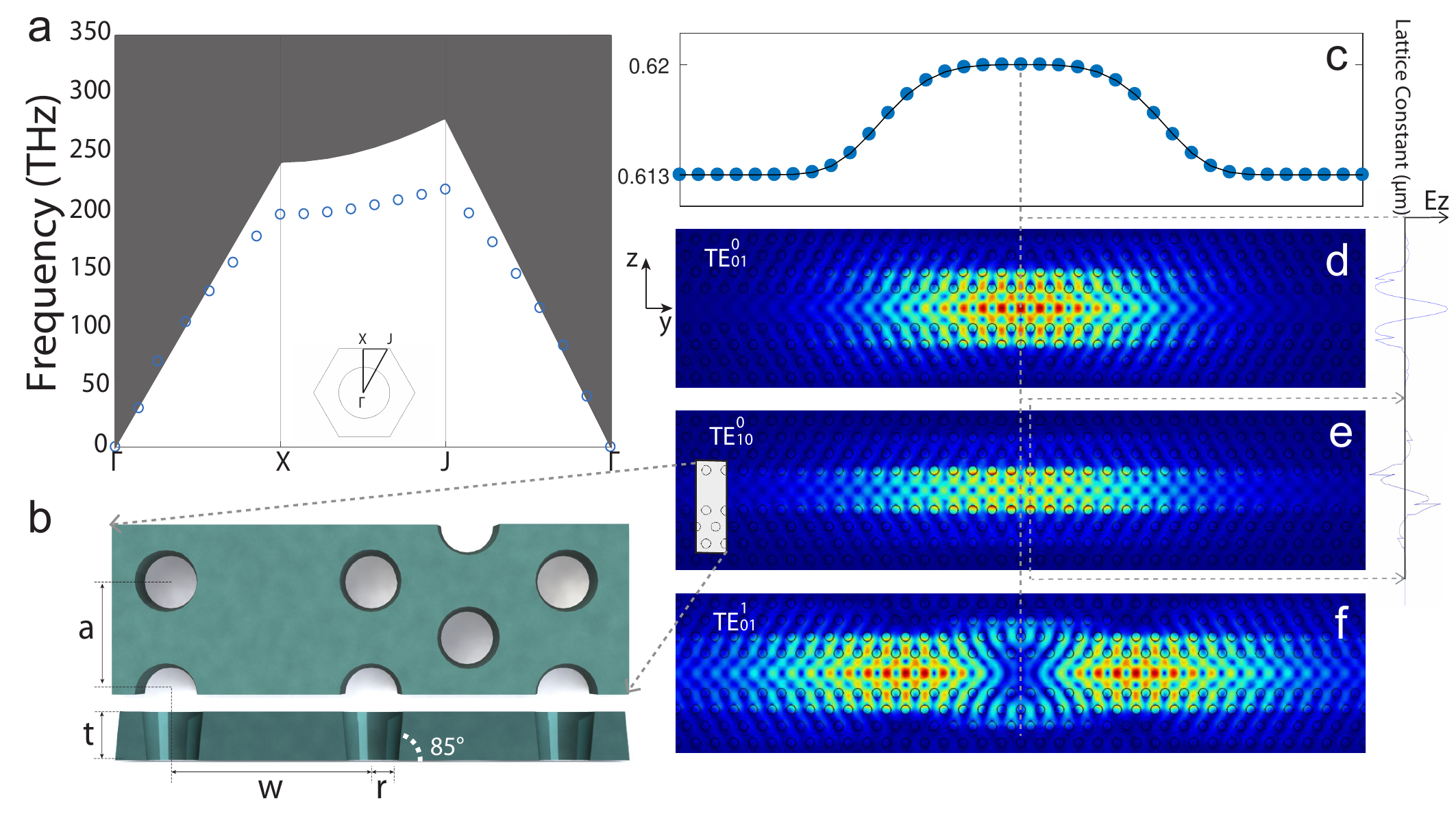}
	\caption{\label{Fig1} Properties of the photonic band structure and line-defect cavity modes of the designed 2D LN photonic crystal slab. \textbf{a.} Dispersion property of the fundamental transverse-electric-like (TE-like) guided mode inside the designed 2D photonic crystal slab. {\bf b.} Schematics showing the top view and cross section of the 2D PhC slab with a line defect waveguide. {\bf c.} Lattice constant as a function of position, which is optimized for high radiation-limited optical Q. {\bf d-f.} The optical mode field profiles of the fundamental (${\rm TE_{01}^0}$ and ${\rm TE_{10}^0}$) and second-order (${\rm TE_{01}^1}$) TE cavity modes, with electric field dominantly lying in the device plane. The mode field profiles are simulated by the finite element method. The left inset shows the orientation of crystal where the optical axis is along the $z$ direction. The right inset shows the ${\rm E_{z}}$ cavity fields as a function of transverse position, at the cross sections indicated by the dashed lines. }
\end{figure*}

In this paper, we demonstrate 2D LN PhC slab nanoresonators with optical Q up to $3.51\times 10^5$, about three orders of magnitude higher than other 2D LN photonic crystal nanocavities reported to date \cite{Baida05, Gu06, Salut06, Gunter09, Pertsch10, Laude10, Baida11, Bernal12, Diziain13, Wang14, Pertsch14}. The high optical Q together with the tiny effective mode volume supports extremely strong nonlinear optical interactions, which results in interesting third harmonic generation, for the first time in on-chip LN nanophotonic devices \cite{Gunter07, Fathpour13, Reano14, Loncar14, Cheng15, Pertsch15, Xu15, Hu15, Shayan16, Jiang16, Bower16, Fathpour17, Loncar17, Luo17, Sun17, Luo172, Amir17, Liang17, Buse17, Loncar173, Peruzzo17, Loncar172}. In particular, the pure polarization characteristics of the cavity modes enabled us to reveal peculiar anisotropy of photorefraction quenching and unique anisotropic thermo-optic nonlinear response, both of which have never been reported before. It also allowed us to observe strong orientation-dependent generation of second harmonic. The demonstrated high-Q 2D LN PhC nanoresonators not only offer an excellent device platform for the exploration of extreme nonlinear and quantum optics at single-photon and few-photon level, but also open up up a great avenue towards future development of energy efficient nonlinear photonic and electro-optic signal processing.

\section*{Device design and fabrication}

\begin{figure*}[htbp]
	\centering\includegraphics[width=2.0\columnwidth]{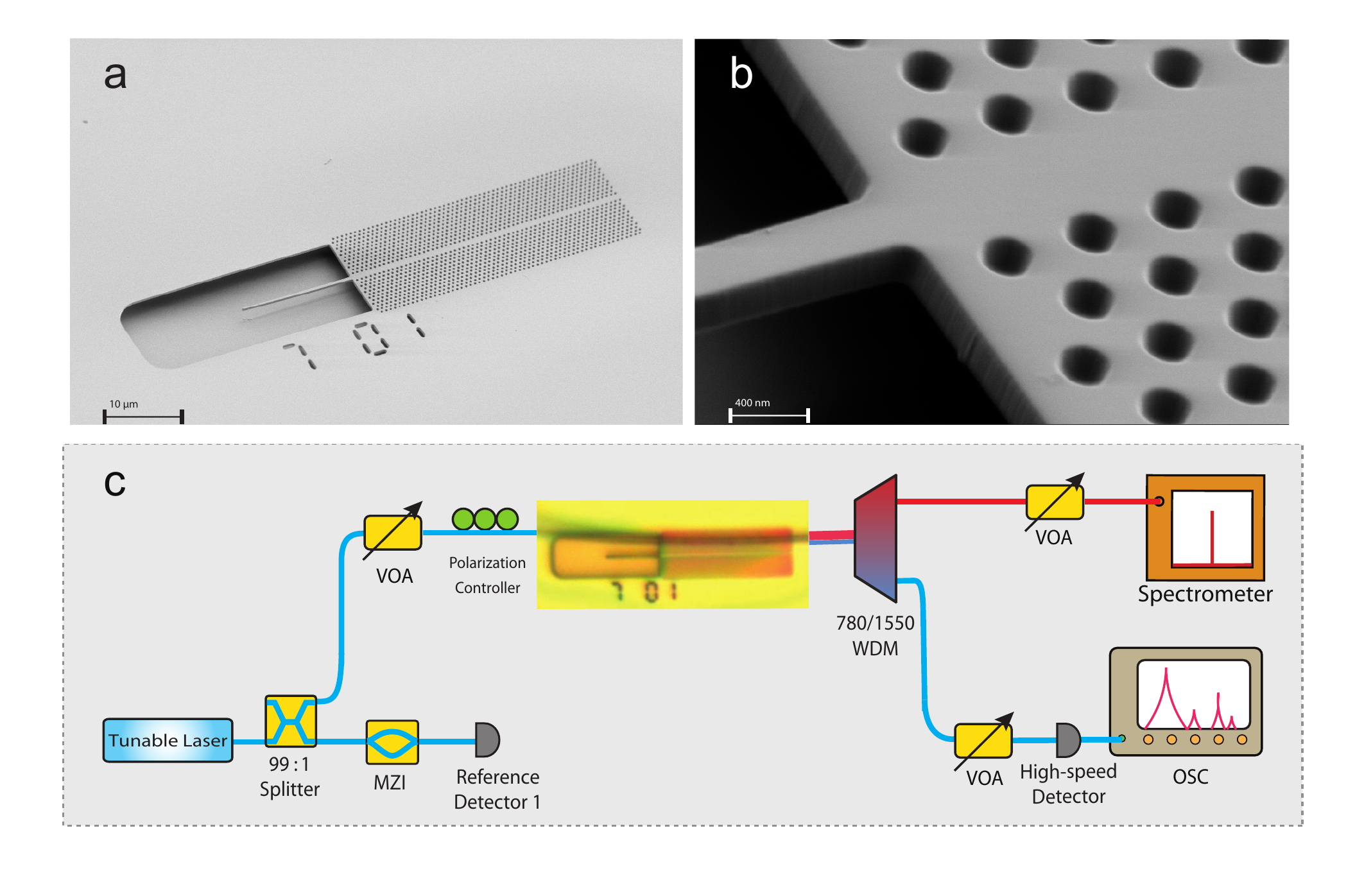}
	\caption{\label{Fig2} Fabricated device structure and experimental testing setup. {\bf a.} Scanning electron microscopic image of a fabricated 2D LN PhC slab. {\bf b.} Zoom-in image of a section of the photonic crystal slab. {\bf c.} Schematic of the experimental testing setup. MZI: Mach-Zehnder interferometer, used to calibrate the laser wavelength; VOA: variable optical attenuator; WDM: wavelength-division multiplexed filter; OSC: oscilloscope. The inset shows an optical microscopic image of a device side coupled to a tapered optical fiber. }
\end{figure*}

The major challenge in making high quality 2D LN photonic crystal structure lies in the complexity of device structures and the stringent requirement on fabrication precision which are significantly beyond current technology to define LN nanophotonic structures. For example, current plasma etching approaches generally produce a slant angle on the device sidewall \cite{Loncar14, Jiang16}. Although such a slant angle can be incorporated into the design of 1D LN photonic crystal \cite{Liang17}, it imposes serious challenge in both the design and fabrication of 2D structures which has much more stringent requirement on the sidewall verticality. On the other hand, LN is a ferroelectric crystal with strong material anisotropy, which leads to significant etching anisotropy in the device plane, making it challenging to define nanoscopic circular hole structures required by 2D photonic crystals \cite{Baida05, Gu06, Salut06, Gunter09, Pertsch10, Laude10, Baida11, Bernal12, Diziain13, Wang14, Pertsch14}.

We overcame all these challenges by optimizing the fabrication processes, which were able to produce a sidewall angle of about ${\rm 85^o}$ uniformly across the whole device structure (Fig.~\ref{Fig1} and \ref{Fig2}). Such a nearly vertical device sidewall enabled us to design well-defined photonic guided modes, as shown in Fig.~\ref{Fig1}a. We optimized the layer thickness $t$, hole radius $r$, and lattice constant $a$ of a hexagonal structure to enlarge the photonic bandgap. Detailed simulations by the finite element method shows that a layer thickness of 270~nm, a hole radius of 145~nm, together with a lattice constant of 620~nm is able to achieve an optimized bandgap of 28.1 THz in the telecom band, with the fundamental transverse-electric (TE) polarized guided mode well confined below the light line (Fig.~\ref{Fig1}a).

To produce a well-confined line-defect cavity, we employed the multi-heterostructure design \cite{Noda05}, by gradually varying the lattice constant from 620~nm to 613 nm around the center of the line-defect waveguide (Fig.~\ref{Fig1}b and c) and optimizing the waveguide width $w$ to minimize the radiation leakage. Figure \ref{Fig1}d-e show the simulated optical mode field profiles of the fundamental (${\rm TE_{01}^0}$ and ${\rm TE_{10}^0}$) and second-order (${\rm TE_{01}^1}$) TE-like cavity modes, which exhibit radiation-limited optical Qs of $1.5 \times 10^6$, $5 \times 10^5$, and $3 \times 10^5$, respectively, with effective mode volumes of 2.43$(\lambda/n)^3$, 3.06$(\lambda/n)^3$, and 4.63$(\lambda/n)^3$. In particular, the nearly vertical device sidewalls here significantly decrease the polarization hybridization, in contrast to the 1D photonic crystal nanobeams demonstrated recently \cite{Liang17}. For example, the fundamental cavity mode ${\rm TE_{01}^0}$ shown in Fig.~\ref{Fig1}d exhibits 62.5\% of its energy in the $z$-polarization lying in the device plane, and almost zero in the $x$-direction. Such a pure polarization enables us to explore intriguing anisotropy of optical phenomena, by making the line-defect cavity either in parallel with or perpendicular to the optical axis, as we will show below. For convenience, we denote the one perpendicular to the optical axis as an \emph{e-cavity} since the dominant electric field polarizes along the optical axis, corresponding to the extraordinary polarization (see Fig.~\ref{Fig1}d). Accordingly, we denote the one in parallel with the optical axis as an \emph{o-cavity} as the dominant cavity field polarizes along the ordinary polarization.

Our devices were fabricated on a $270$-nm-thick X-cut single-crystalline LN thin film sitting on a 2-${\rm \mu m}$ silicon dioxide layer on a silicon substrate. To define the photonic crystal structure, we deposited first a $400$-nm thick amorphous silicon as a hard-etching mask through plasma enhanced chemical vapor deposition. The device structure was patterned with ZEP-$520$A positive resist via electron-beam lithography, which was then transferred to the amorphous silicon mask layer with a standard reactive ion etching process. It was in turn transferred to the LN layer with an Ar$^+$ plasma etching process. The residual mask was removed by a $30$\% KOH resolvent at 70 $^{\rm o}$C. Finally, we use diluted hydrofluoric acid to undercut the buried oxide layer to form a suspended photonic crystal membrane structure (Fig.~\ref{Fig2}a and b). 

\section*{Optical properties}

Figure \ref{Fig2}a and b show an example of a fabricated device. It shows clearly that we achieved a nearly vertical device sidewall both inside and outside the circular holes, uniformly across the entire device structure. The circular shape of holes is well defined and the hexagonal lattice structures is accurately patterned. To characterize the optical property of the device, we launched a continuous-wave tunable laser into the device via evanescent side-coupling with a tapered optical fiber. Figure \ref{Fig2}c shows the schematic of the experimental testing setup, where the optical wave transmitted out from the device is detected by a photodetector whose output is characterized by an oscilloscope. In the case of harmonic generation, the harmonic light was separated by a dichroic filter and then recorded by a spectrometer.

\begin{figure*}[t!]
	\centering \includegraphics[width=1.550\columnwidth]{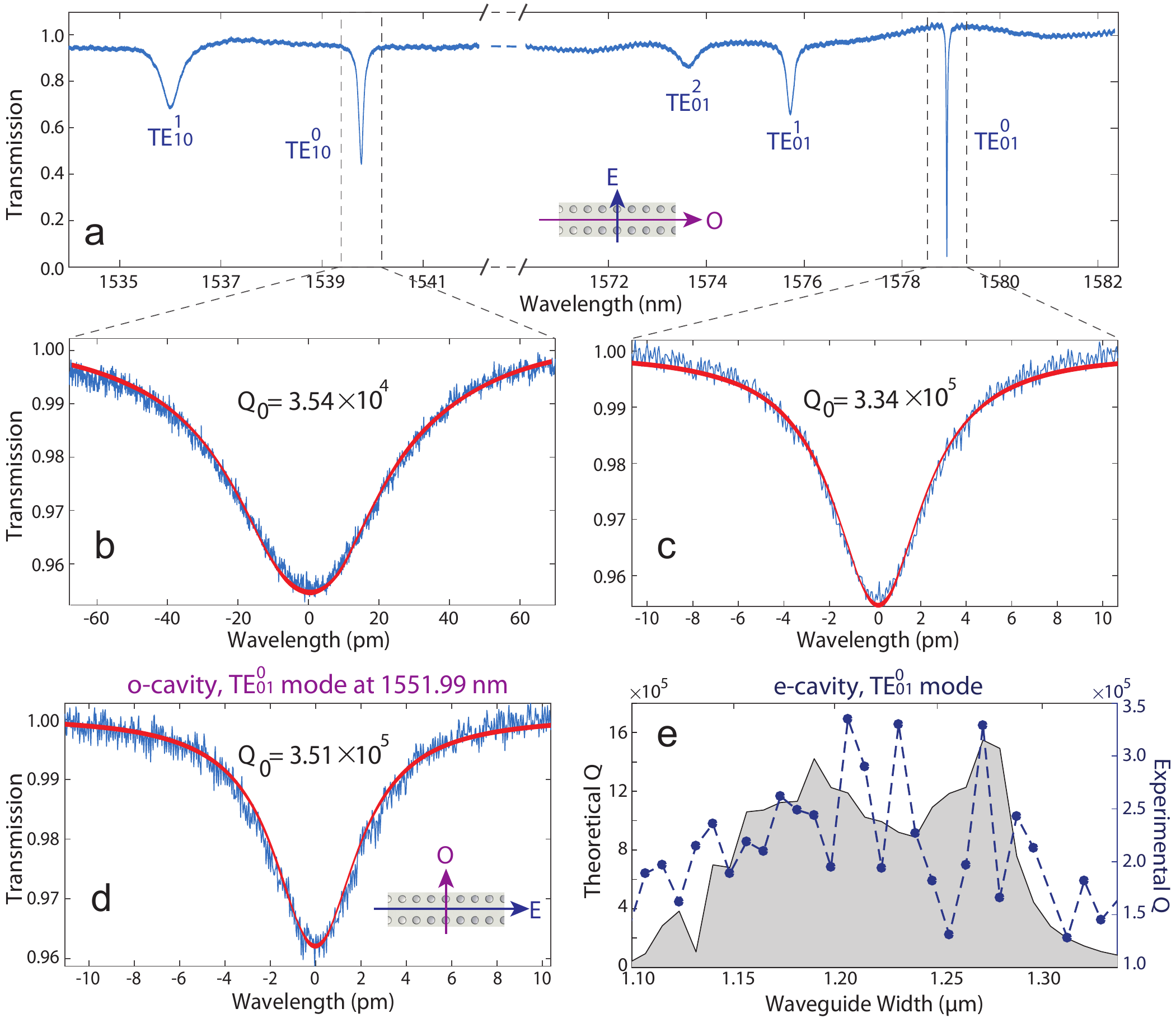}
	\caption{\label{Fig3} Linear optical properties of 2D LN PhC slab nanoresonators. {\bf a.} Laser-scanned transmission spectrum of an e-cavity. The inset shows the top-view schematic of the e-cavity device orientation with respect to the optical axes. {\bf b and c.} Detailed transmission spectra of the ${\rm TE_{10}^0}$ and ${\rm TE_{01}^0}$ cavity modes, respectively, with the experimental data shown in blue and the theoretical fitting shown in red. {\bf d.} Detailed transmission spectrum of the fundamental ${\rm TE_{01}^0}$ cavity mode in an o-cavity, with the experimental data shown in blue and the theoretical fitting shown in red. The inset shows the top-view schematic of the o-cavity device orientation. {\bf e.} Optical Qs of the ${\rm TE_{01}^0}$ mode in e-cavities as a function of the width of line-defect waveguide. The blue dots show the experimental data and the solid line shows the theoretical results simulated by the finite element method. The fluctuations on the theoretical curve are primarily due to the precision of numeric simulations, which is limited by the finite size of the computer memory. The dashed line is used for eye guidance only. }
\end{figure*}

Figure \ref{Fig3}a shows the transmission spectrum of an e-cavity device recorded when we scanned the laser over a telecom band. It shows that the devices exhibit five cavity modes located at two spectral regions between 1535 and 1541~nm and between 1572 and 1582~nm, with the mode species identified on the figure. Detailed characterizations (Fig.~\ref{Fig3}b and c) show that the fundamental cavity mode ${\rm TE_{ 01}^0}$ located at 1578.96~nm exhibits an intrinsic optical Q as high as $3.34 \times 10^5$, while the second-order mode ${\rm TE_{10}^0}$ at 1539.36~nm has an optical Q about one order of magnitude lower. The o-cavity devices exhibit same magnitude of optical quality. One example is shown in Fig.~\ref{Fig3}d, which shows an optical Q of $3.51 \times 10^5$ that is even higher than that of the e-cavity. These optical Qs are about three orders of magnitude higher than other 2D LN photonic crystal nanocavities reported to date \cite{Baida05, Gu06, Salut06, Gunter09, Pertsch10, Laude10, Baida11, Bernal12, Diziain13, Wang14, Pertsch14}. They are only about five times lower than the theoretically designed values, indicating the high quality of device fabrication. They are even more than three times higher than the 1D LN nanoresonators we demonstrated very recently \cite{Liang17}, although the 2D device structure here is significantly more complicated.

To obtain the dependence of optical Q on the width of line-defect waveguide, we characterized more than 50 e-cavity nanoresonators with varied waveguide width. Figure \ref{Fig3}e shows the results for the ${\rm TE_{01}^0}$ mode. It shows clearly that the experimentally recorded optical Q maintains above $1.5 \times 10^5$ for a large range of waveguide width between 1.10 and 1.33~${\rm \mu m}$, with a peak value appearing at $w= 1.27$ $\mu$m. This trend agrees well with the numerical simulations shown in the solid line. The o-cavities exhibit similar dependence on the waveguide width, while the waveguide width leading to the peak optical Q shifts to $w = 1.18$ ${\rm \mu m}$, simply due to the birefringence of LN crystal.

\section*{Photorefraction}

\begin{figure*}[t!]
	\centering\includegraphics[width=2.0\columnwidth]{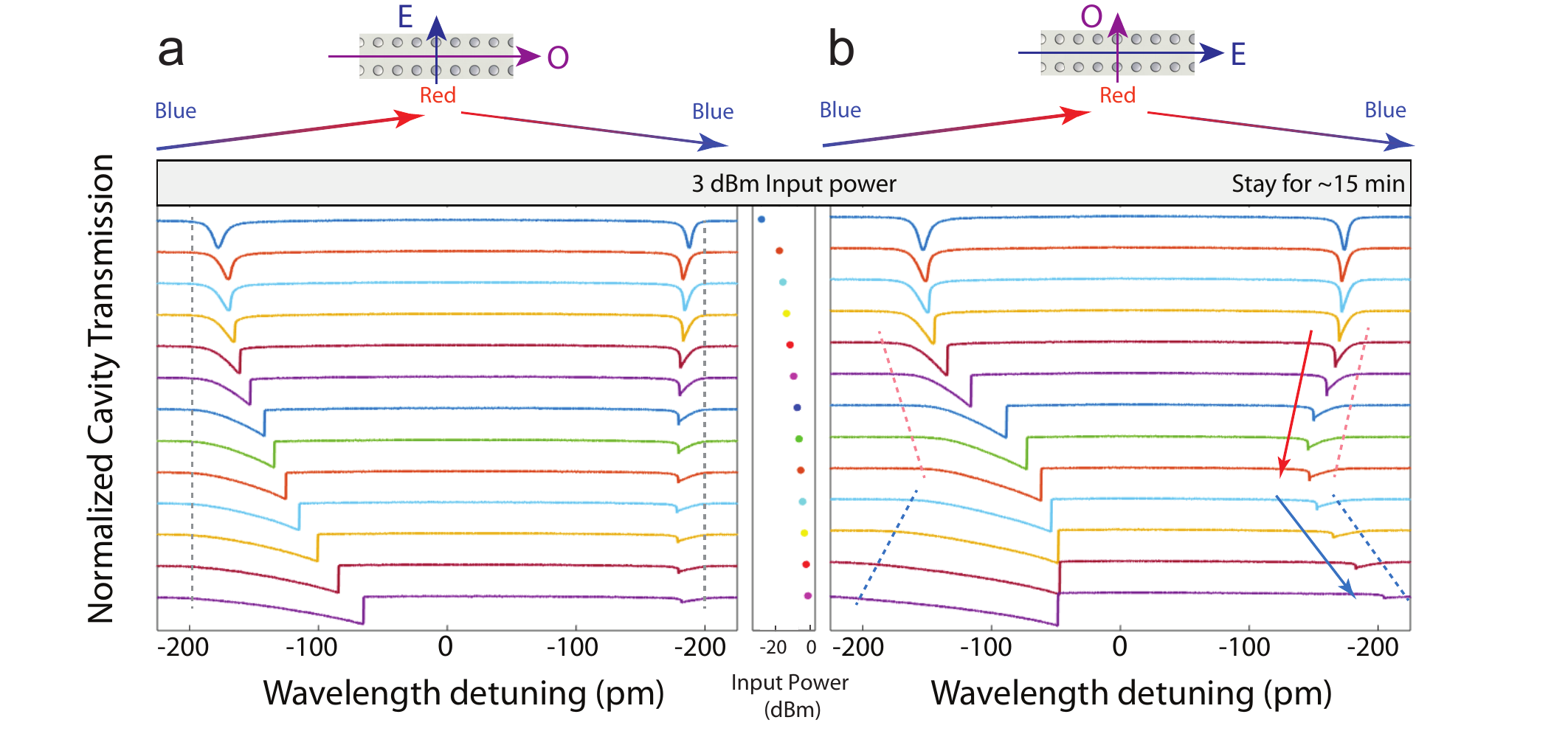}
	\caption{\label{Fig4} Laser-scanned cavity transmission spectrum as a function of input power. The input optical power remains first at 2 mW for about 15 minutes (gray region) to quench the photorefraction before we started recording the transmission spectra at different power levels. The input power corresponding to each scanning spectrum is shown in the middle. The laser wavelength is periodically scanned back and forth in a triangular fashion over a spectral range of 230~pm, with a scanning period of 100 ms. The cavity transmission spectra are shifted with each other along the vertical axis for convenient comparison. The left panel is recorded for an e-cavity (indicated on the top), where the dashed line indicates the left edge of the cavity resonance which remains unchanged with optical power. The right panel is for an o-cavity, where the red and blue dashed lines show the red and blue shifts, respectively, of the left edge of the cavity resonance. }
\end{figure*}

The high quality and the small effective mode volume of the 2D LN photonic crystal slab nanoresonators would result in dramatic enhancement of optical energy density inside the cavity, which would thus support strong nonlinear optical interaction. In particular, the pure polarization of the cavity modes in these devices enables us to explore the potential anisotropy of nonlinear optical phenomena. This anisotropy is challenging to access in 1D LN photonic crystal nanobeams \cite{Liang17} that exhibit significant polarization hybridization due to the specific shape of the device cross section. In the following, we explore photorefractive effect, thermo-optic bistability, and harmonic generation, which show intriguing behaviors in the nanoscopic scale that do not appear in any other LN photonic devices.

To do so, we selected an e-cavity and an o-cavity with the same intrinsic optical Q of $3.3 \times 10^5$ for the fundamental ${\rm TE_{01}^0}$ cavity mode. For a fair comparison between the two devices, we maintained exactly same operation conditions for the two devices, by positioning the coupling tapered fiber to maintain a same external coupling efficiency of $70$\% for both devices (accordingly, same loaded optical Q around $2.1 \times 10^5$ for both cavities). To explore the photorefractive effect, we increased the input optical power to 2 mW, at which the photorefraction is quenched in both devices, an intriguing phenomena we also observed previously in the 1D LN photonic crystal nanobeams \cite{Liang17}. We maintained the devices at this condition for about 15 minutes to stabilize the photorefraction quenching. After that, we varied the optical power and monitored the transmission of the devices when we continuously scanned the laser wavelength across the cavity resonances back and forth in a periodic triangular fashion.

Figure \ref{Fig4} shows the laser-scanned transmission spectra of the cavities at different optical power levels. The quenching of photorefraction behaves very differently in the two cavities. In the e-cavity, Fig.~\ref{Fig4}a shows that the left edge of the cavity resonance remains unchanged when the input power is varied between the whole range from 2 $\mu$W to 0.8 mW, as indicated by the dashed line. Since the left edge of the cavity resonance indicates the spectral location of the passive cavity (in the absence of optical wave), this implies a complete quenching of the photorefraction. In the o-cavity, surprisingly, Fig.~\ref{Fig4}b shows that the left edge of the cavity resonance shifts towards red when the input power increases from 0.08 mW to 0.32 mW, as indicated by the red dashed line. However, it shifts backwards towards blue when the optical power increases further from 0.32 mW to 0.8 mW, as indicated by the blue dashed line. The whole process is reversible when the optical power changes. To the best of our knowledge, this is the first time to observe such peculiar anisotropic behavior of photorefraction quenching. The underlying mechanism is not clear at this moment. One potential reason is likely related to the photovoltaic property of lithium niobate, which tends to produce photocurrent preferably along certain crystographic axis \cite{Gaylord85}. However, the exact physical nature requires further exploration in the future.

\section*{Thermo-optic nonlinearity}
\begin{figure*}[t!]
	\centering\includegraphics[width=2.0\columnwidth]{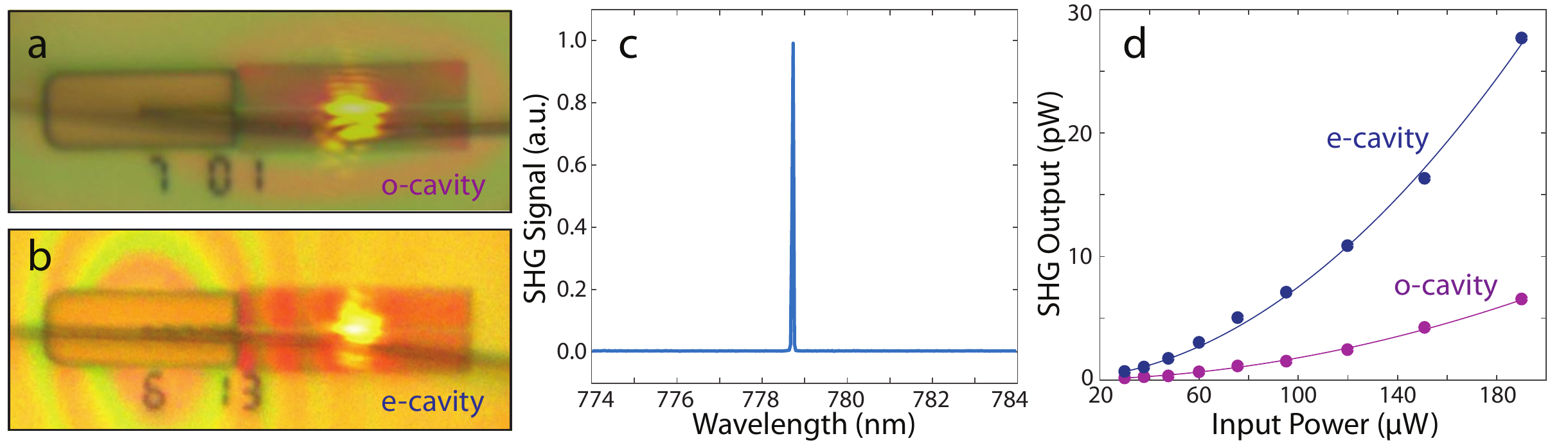}
	\caption{\label{Fig5} Second harmonic generation in 2D LN PhC slab nanoresonators. {\bf a and b.} Optical microscopic images of an e-cavity and o-cavity, respectively, showing bright spots appearing when the pump wave is launched into the cavities. {\bf c.} Spectrum of generated second harmonic, with a pump wave at 1557.45 nm. {\bf d.} Recorded power of the generated second harmonic wave as a function of that of the fundamental pump wave, where the blue and purple dots are experimental recorded data for an e-cavity and o-cavity, respectively. The solid lines show quadratic fitting to the experimental data. }
\end{figure*}
Figure \ref{Fig4}a and b show clear thermo-optic bistability when the input optical power becomes significant, as expected. However, it is interesting to note that, at a same power level, the thermo-optic bistability in the o-cavity is considerably larger than that in the e-cavity. This is surprising since the two devices exhibit identical intrinsic optical Q and operates with exactly same external coupling condition, from which we expect a same temperature change due to photothermal heating. As the thermo-optic coefficient for the extraordinary polarization ($\frac{dn_e}{dT} = 3.34 \times 10^{-5}{\rm /K}$) is significantly greater than that for the ordinary light ($\frac{dn_o}{dT} = 0{\rm /K}$) at room temperature in the telecom band, and the thermal expansion along the optical axis ($\alpha^{(z)} = 0.75 \times 10^{-5} {\rm /K}$) is only slightly smaller than that along the orthogonal direction ($\alpha^{(x,y)} = 1.54 \times 10^{-5} {\rm /K}$)), we shall expect a larger thermo-optic bistability in the e-cavity instead.

The peculiar thermo-optic nonlinear behavior is likely due to the pyroelectricity of lithium niobate \cite{Gaylord85}, which produces an electric field along the crystal axis when temperature increases. The induced electric field in turn decreases the refractive index via the Pockels effect, which compensates the refractive index increase induced by the thermo-optic effect. As the Pockels effect is dominant along the optical axis, the e-cavity will experience more the pyroelectricity-induced refractive index change, leading to a smaller net increase of refractive index. Note that the time response of the pyroelectricity is primarily determined by that of the temperature variation which is the same as a normal thermo-optic effect. Consequently, the combined thermo-optic nonlinearity and the pyroelectricity-induced effect would manifest as a net thermo-optic bistability, which is smaller in an e-cavity than an o-cavity, as we observed in Fig.~\ref{Fig4}.

\section*{Harmonic generation}
\begin{figure*}[htpb]
	\centering\includegraphics[width=2.0\columnwidth]{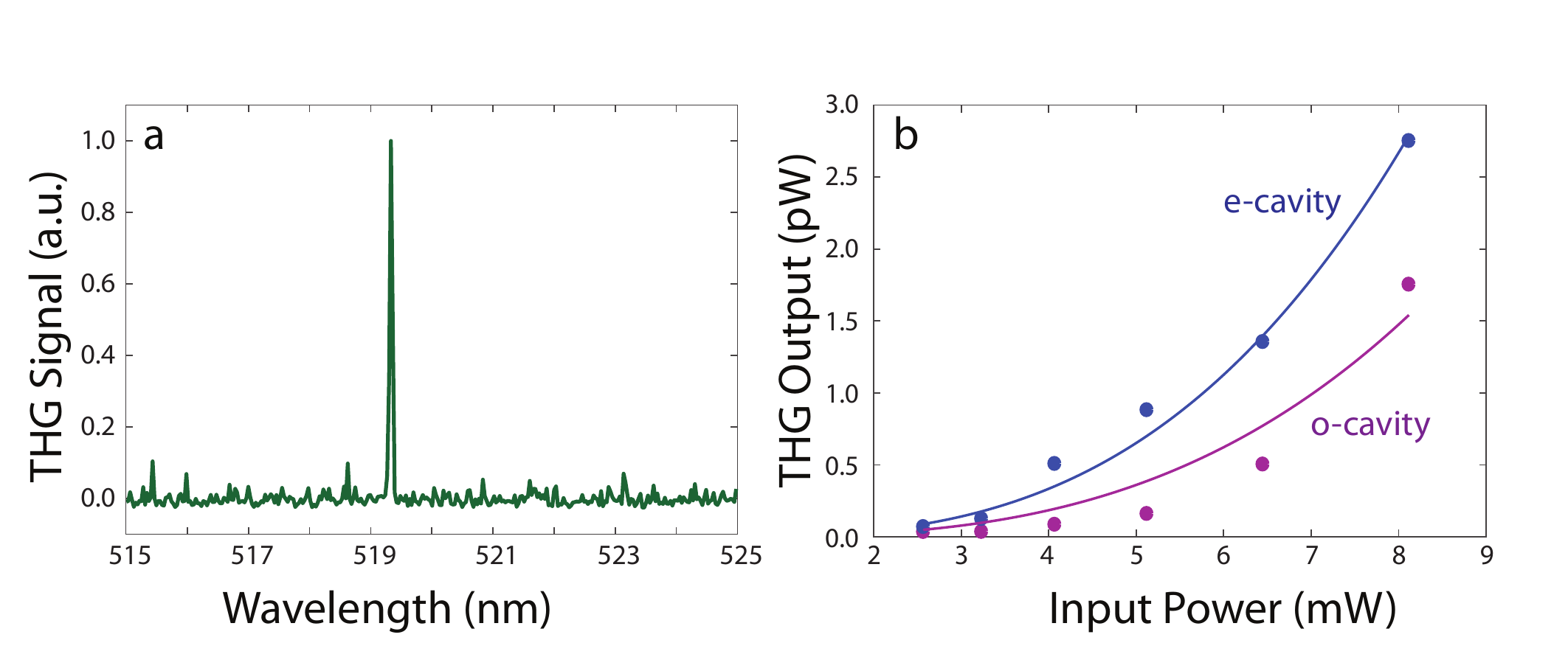}
	\caption{\label{Fig6} Third-harmonic generation in 2D LN PhC slab nanoresonators. {\bf a.} Spectrum of the generated third harmonic, with a pump wave at 1557.45 nm. {\bf b.} Recorded power of the generated third harmonic wave as a function of the fundamental pump wave, where the blue and purple dots are experimental recorded data for an e-cavity and o-cavity, respectively. The solid lines show cubic fitting to the experimental data. }
\end{figure*}
Lithium niobate exhibits significant second-order optical nonlinearity. Therefore, the strong nonlinear optical interactions in the 2D LN PhC slab nanoresonators would enable us to observe second-harmonic generation (SHG). Figure \ref{Fig5} shows an example. When we increased the optical power dropped into the e-cavity, a bright spot appeared at the center of the device structure where the nanoresonator is located, which is clearly visible in the optical microscopic image (Fig.\ref{Fig5}a). As the imaging camera has a spectral response in the visible and near infrared, the bright spot implies the generation of second harmonic. This is verified by the emission spectrum shown in Fig.~\ref{Fig5}c which shows a clear coherent emission line at 778.73 nm when we launched a pump wave into the cavity mode at 1557.45 nm. The SHG exhibits a quadratic power dependence, as shown in Fig.~\ref{Fig5}d, which is a typical signature of second harmonic generation. Figure \ref{Fig5}d shows a conversion efficiency of 0.078~${\rm \%/W}$. The small value is primarily due to the SHG frequency which is well above the light line of the 2D photonic crystal (Fig.~\ref{Fig1}a) that was designed only for high-Q cavities in the telecom band, leading to low optical Q around the second harmonic with significant radiation leakage into free space. On the other hand, the coupling tapered optical fiber was designed for operating in the telecom band, which exhibits very low coupling efficiency at shorter waveband. Future optimization of the device design and the coupling waveguide would help improve the nonlinear conversion efficiency.

We also characterized the second harmonic generation in an o-cavity, as shown in Fig.~\ref{Fig5}d. Compared with the e-cavity, the o-cavity exhibits an efficiency about 4 times lower. It is possibly because the SHG in the e-cavity is likely to be a type-0 process since the dominant cavity field polarizes along the optical axis where the nonlinear susceptibility is maximum, while that in the o-cavity is likely to be a type-II process since the cavity field is dominantly along the ordinary polarization. Interestingly, Fig.~\ref{Fig5}a and b show that SHG in the e-cavity tends to be spatially located towards the side of the line-defect cavity, while that in the o-cavity tends to be located at the center. The exact physical reason is not clear at this moment, which requires further exploration.

Of particular interest is that the nonlinear optical interaction is enhanced so much in the 2D LN PhC slab nanoresonators that we were even able to observe third-harmonic generation. This is shown in Fig.~\ref{Fig6}, where a clear emission line appears at a wavelength of 519.15 nm, which corresponds directly to the third harmonic of the pump wave at 1557.45 nm. We recorded the power dependence of the third harmonic, which is plotted in Fig.~\ref{Fig6}b. It shows a clear cubic dependence on the pump power, an intrinsic signature of third harmonic generation. To the best of our knowledge, this is the first time to observe third-harmonic generation in on-chip LN nanophotonic devices \cite{Gunter07, Fathpour13, Reano14, Loncar14, Cheng15, Pertsch15, Xu15, Hu15, Shayan16, Jiang16, Bower16, Fathpour17, Loncar17, Luo17, Sun17, Luo172, Amir17, Liang17, Buse17, Loncar173, Peruzzo17, Loncar172}. Similar to the case of the second harmonic generation, the third-harmonic generation is considerably weaker in the o-cavity (Fig.~\ref{Fig6}b, purple curve), which is likely due to the tensorial nature of the third-order nonlinear susceptibility of lithium niobate.

\section*{Discussion}

We have demonstrated 2D LN PhC slab nanoresonators with optical Q up to $3.51 \times 10^5$ that is about three orders of magnitude higher than other 2D LN PhC nanoresonators reported to date \cite{Baida05, Gu06, Salut06, Gunter09, Pertsch10, Laude10, Baida11, Bernal12, Diziain13, Wang14, Pertsch14}. The high optical Q together with tight optical mode confinement results in intriguing nonlinear optical interactions. We have observed second-harmonic generation, particularly third harmonic generation that is the first time to be observed in on-chip LN nanophtonic devices \cite{Gunter07, Fathpour13, Reano14, Loncar14, Cheng15, Pertsch15, Xu15, Hu15, Shayan16, Jiang16, Bower16, Fathpour17, Loncar17, Luo17, Sun17, Luo172, Amir17, Liang17, Buse17, Loncar173, Peruzzo17, Loncar172}. Moreover, the devices exhibits pure polarization of the cavity modes, which enabled us to probe the intriguing anisotropy of nonlinear optical phenomena. We have revealed the peculiar anisotropy of photorefraction quenching and the unique anisotropic thermo-optic nonlinear response, which have never been reported previously.

With the significant optical nonlinearity of lithium niobate material, the demonstrated high-Q 2D LN PhC nanoresonators offer an excellent device platform for exploring extreme nonlinear and quantum optical phenomena in the regime inaccesible to conventional means. For example, the low nonlinear conversion efficiency in our current devices is primarily due to the low optical Q and external coupling at the second and third harmonic. Therefore, future design to minimize the radiation leakage at the second (and/or third) harmonic frequency would significantly improve the optical Q. Moreover, appropriate design of on-chip coupling waveguide would optimize the external coupling to these waveband. The resulting device platform would further dramatically enhance nonlinear wave interactions, which would enable exploration of nonlinear and quantum optics at single and few photon levels, which would open up broad nonlinear and quantum photonic applications \cite{Lukin14}.

On the other hand, 2D PhC slab resonators are particularly suitable for optoelectronic integration \cite{Painter11, Baba12} and for scaling up to large-scale photonic integrated circuits \cite{Notomi14}. As lithium niobate exhibits strong piezoelectric effect, electro-optic effect, and electromechanical coupling, the demonstration of high-Q 2D LN PhC slab nanoresonators thus open the door towards developing large-scale electro-opto-mechanically hybrid integrated LN circuits for broad applications in communication, computing, signal processing, sensing, and energy harvesting.

\noindent\textbf{Acknowledgements} This work was supported in part by the National Science Foundation under Grant No.~ECCS-1641099 and ECCS-1509749, and by the Defense Advanced Research Projects Agency SCOUT program through Grant No. W31P4Q-15-1-0007 from the U.S. Army Aviation and Missile Research, Development, and Engineering Center (AMRDEC). The views and conclusions contained in this document are those of the authors and should not be interpreted as representing the official policies, either expressed or implied, of the Defense Advanced Research Projects Agency, the U.S. Army, or the U.S. Government. This study was performed in part at the Cornell NanoScale Science and Technology Facility (CNF), a member of the National Nanotechnology Infrastructure Network.

\end{document}